%Paper: hep-th/9309119
%From: poly@dxcern.cern.ch (Alexios Polychronakos)
%Date: Wed, 22 Sep 1993 12:34:38 +0200

%%%%%%%%%%%%%%%%%%%%%%%%%%%%%%%%%%%%%%%%%%%%%%%%%%%%%%%%%%%%%%%%%%%%%%%%%%%%%%%
%%%%% This Tex file inputs phyzzx.       				  %%%%%
%%%%%%%%%%%%%%%%%%%%%%%%%%%%%%%%%%%%%%%%%%%%%%%%%%%%%%%%%%%%%%%%%%%%%%%%%%%%%%%
\input phyzzx
\def\CR{C_{2R}}

\def\be{\beta}
\def\al{\alpha}
\def\De{\Delta}

\def\tu{\widetilde u}
\def\tQ{\widetilde Q}
\def\th{\theta}

\def\nt{\tilde n}
\def\izl{\int_0^L dx}
\def\Ad{\dot A_1}
\def\Wd{\dot W}

\def\lam{\lambda}
\def\de{\delta}
\def\eps{\epsilon}

\def\ws{world-sheet}
\def\QCD{${\rm QCD}_2$}

\def\cZ{{\cal Z}}

\def\im{{\rm Im}}
\def\re{{\rm re}}

\Pubnum={CERN-TH-7016/93\cr
UVA-HET-93-08\cr
hepth@xxx/9309119}
\date={September 1993}
\titlepage
\title{Classical Solutions for Two Dimensional QCD on the Sphere}
\bigskip
\author {
Joseph~A.~Minahan\footnote\star
{Present address: Department of Physics, University of Southern California,
Los Angeles, CA 90089-0484; minahan@physics.usc.edu}}
\address{Department of Physics,  Jesse Beams Laboratory,\break
University of Virginia, Charlottesville, VA 22901 USA}
\andauthor{
Alexios P. Polychronakos\footnote\dagger
{poly@dxcern.cern.ch}}
\address{Theory Division, CERN\break
CH-1211, Geneva 23, Switzerland}
\bigskip
\abstract{
We consider $U(N)$  and $SU(N)$ gauge theory on the sphere. We express
the problem in terms of a matrix element of $N$ free fermions on a circle.
This allows us to find an alternative way to show Witten's result
that the partition function is a sum over classical saddle points.
We then show how the phase transition of Douglas and Kazakov occurs from
this point of view. By generalizing the work of Douglas and Kazakov, we
find other ``stringy" solutions for the $U(N)$ case in the large $N$ limit.
Each solution is described by a net $U(1)$ charge. We derive a relation for
the maximum charge for a given area and we also describe the critical behavior
for these new solutions. Finally, we describe solutions for lattice $SU(N)$
which are in a sense dual to the continuum $U(N)$ solutions.
}
%\submit{}
\vfill
\endpage
\def\NP{{\it Nucl. Phys.\ }}
\def\PL{{\it Phys. Lett.\ }}
\def\PRD{{\it Phys. Rev. D\ }}

\def\MPL{{\it Mod. Phys. Lett. A\ }}

\def\ZETF{{\it Zh. Eksp. Teor. Fiz.}}

\REF\GW{D. Gross and E. Witten, \PRD {\bf 21} (1980) 446.}
\REF\Wadia{S. Wadia, \PL {\bf 93B} (1980) 403.}
%\REF\tHooft{G.~'t~Hooft, \NP {\bf B75} (1974) 461.}
%\REF\BBHP{W.~Bardeen, I.~Bars, A.~Hanson and R.~Peccei, \PRD {\bf13} (1976)
%2364.}
%\REF\BarsI{I.~Bars, \PRL {\bf36} (1976) 1521; \NP {\bf B111} (1976) 413.}
\REF\Gross{D.~Gross, \NP {\bf B400} (1993) 161.}
\REF\JM{J.~Minahan, \PRD {\bf 47} (1993) 3430.}
\REF\GrTay{D.~Gross and W.~Taylor, \NP {\bf B400} (1993) 181.}
\REF\GrTayII{D.~Gross and W.~Taylor, CERN-TH-6827-93 hep-th/9303046, 1993.}
\REF\DK{M. Douglas and V. Kazakov, hep-th/9305047.}
\REF\Witten{E. Witten, {\it J. Phys.} {\bf G9} (1992) 303.}
\REF\MP{J. Minahan and A. Polychronakos, \PL {\bf B312} (1993) 155.}
\REF\Mike{M. Douglas, hep-th/9303159, 1993.}
\REF\ROI{M. Caselle, A. D'Adda, L. Magnea, and S. Panzeri, hep-th/9304015,
1993.}
\REF\Migdal{A.~Migdal, \ZETF {\bf 69} (1975) 810.}
\REF\Rus{B.~Rusakov, \MPL {\bf5} (1990) 693.}
\REF\DH{J. Duistermaat and G. Heckman, {\it Invent. Math} {\bf 69} (1982) 259.}
\REF\APA{A.P.~Polychronakos, \PL {\bf B266} (1991) 29.}
%\REF\APG{A.P.~Polychronakos, \PL {\bf B264} (1991) 362.}
\REF\Byrd{P. Byrd and M. Friedman, ``Handbook of Elliptic Integrals for
Engineers and Physicists'', Berlin, Springer, 1954.}
\REF\CDMP{M. Caselle, A. D'Adda, L. Magnea, and S. Panzeri,
hep-th/9309107, 1993.}

\chapter{Introduction}

In their classic work, Gross and Witten[\GW], and independently Wadia[\Wadia],
showed
that lattice QCD in two dimensions contains a third order phase transition
in the large $N$ limit.  Basically, what happens is that the partition function
for the plaquettes reduces to a product of partition functions for the
individual plaquettes.  The
theory on each one is equivalent to a $d=0$ unitary matrix model,
with the potential given by
$$V={N\over g^2a^2}\tr(U+U^\dagger).\eqn\matpot$$
This theory is equivalent to zero-dimensional field theory of $N$ eigenvalues
with values that lie on a circle and a
potential
$$V=-(N/2g^2a^2)\sum_i(e^{i\th_i}+e^{-i\th_i})
-\sum_{i<j}\log\left(\sin{\th_i-\th_j\over2}\right)^2.\eqn\Vpot$$
The theory becomes critical by tuning the lattice size such that the highest
fermion of the Dyson gas lies at the top of the potential.  For fixed
$g$, as the lattice size is decreased, the potential becomes very deep
and the fermions move further away from the critical configuration.
Thus in the limit of vanishingly small $a$, the fermions should only
see a quadratic potential.

Very recently there has been a revived interest in large $N$ \QCD.  This
program was started by Gross[\Gross], who was looking for a string formulation
of the problem.  Further work in this direction was done by one of
the authors[\JM] and finally, the complete theory was worked out by Gross
and Taylor, who showed that \QCD\ is essentially a string theory for
a target space with arbitrary genus.  The basic insight in handling
this problem is to
equate $U(N)$ or $SU(N)$ representations with sums of maps of
world-sheets into the target space.  It turns out that the number
of boxes for a Young tableau of a representation is equal to the number
of coverings of the surface.  For surfaces with genus $g>1$, organizing
the partition function by the number of covers also nicely organizes
the sum into powers of $1/N$.  Hence it is very easy to compute the
leading order contribution in the large $N$ limit for these surfaces.
For the torus, there can be leading order contributions from any number
of coverings, but it is easy to sum their contribution.
However, in the
case of the sphere finding the leading order contribution in $1/N$
is not so easy, namely because all $SU(N)$ or $U(N)$ representations
contribute to the leading order behavior, and computing these contributions
involves computing $1/N$ corrections to the dimensions of these
representations.

Recently, Douglas and Kazakov (DK) discovered a clever way to solve
the problem of the sphere by treating the rows of the young tableau
and the number of boxes in each row as continuous variables[\DK].  Doing
this they were able to compute the leading behavior which led them
to a surprising result:  the theory contains
a third order phase transition, similar to  the case of the
lattice version of Gross, Witten
and Wadia, but for the continuum limit of the theory.

However, the $U(N)$ case has an interesting feature--- the sum
over representations is not asymptotic.  In order to correct for this,
one must sum over an infinite number of charge sectors.  While this
will eventually lead to overcounting for finite $N$, the answer will
at least be asymptotic in the large $N$ limit.  These different charge
sectors can be thought of as being extra solutions to the large $N$
equations of motion.  These different sectors are conjugate to the $U(1)$
instantons.  Because of this, one should be able to find
the such solutions using the DK analysis.
In this paper we do precisely this by generalizing an ansatz described
by DK.

There has also been some recent work by Witten in \QCD, although from
a different perspective[\Witten].  He has shown that the partition function on
any Riemann surface is given by a sum over the saddle points.  An interesting
question is how the DK phase transition appears from this point of view.

In section two we review recent work on 2d QCD and construct a free fermion
picture for this theory.  In section three we present an alternative derivation
of Witten's result that the partition function is a sum over saddle points.
We then show how the continuum phase transition occurs in this dual picture
of \QCD.  In section 4 we consider the new solutions
for $U(N)$ \QCD and generalize the analysis of DK for
these solutions.  We consider the two cases where the area is near its
critical value, and when the area is very large.  In section 5 we compare
these solutions to corresponding solutions for lattice $SU(N)$ \QCD.
In section 6 we present our conclusions.  We include an appendix with some
useful equations for elliptic integrals.

\chapter{Review of \QCD}

Let us first review how \QCD\ on a cylinder is the same as a
theory of free fermions by showing that it can be reduced to a one-dimensional
unitary matrix model[\MP-\ROI].
In the gauge $A_0=0$, the Hamiltonian is given as
$$H=\half\izl\tr F_{01}^2=\half\izl\tr \Ad^2\eqn\Hamgauge$$
with the overdot denoting a time derivative.
The $A_0$ equation of motion is now the constraint
$$D_1F_{10}=\partial_1\Ad+ig[A_1,\Ad]=0.\eqn\gconstraint$$
Define a new variable $V(x)$,
$$V(x)=W_0^x\Ad(x)W_x^L,\eqn\Vdef$$
where
$$W_a^b={\rm P}e^{ig\int_a^bdxA_1}.\eqn\Wdef$$
Then \Hamgauge\ can be written as
$$\partial_1V(x)=0,\eqn\Veq$$
so $V(x)$ is a constant.  Thus $V(0)=V(L)$, which implies that
$$[W,\Ad(0)]=0,\eqn\WAcomm$$
where $W\equiv W_0^L$ and we have used the periodicity of $A_1$ in $x$.

{}From the definitions \Vdef\ and \Wdef, we find the relation
$$\Wd=ig \izl W_0^x\Ad(x)W_x^L=ig \izl V(x),\eqn\Wdeq$$
and therefore using \Veq\ and \WAcomm, we derive
$$\Wd=igLW\Ad(0)=igL\Ad(0)W.\eqn\WdeqII$$
\WdeqII\ then implies that
$$[W,\Wd]=0.\eqn\WWdeq$$

Because $V(x)=V(0)$, $\Ad(x)$ satisfies
$$\Ad(x)=W_0^x\Ad(0)W_x^0.\eqn\Adeq$$
Thus, using this relation along with \WdeqII, we can rewrite the Hamiltonian
in \Hamgauge\ as
$$H=-{1\over 2g^2L}\tr(W^{-1}\Wd)^2.\eqn\Hammm$$
If the gauge group is $U(N)$, with the $U(1)$ coupling given by $g/N$, then
\Hammm\ is the Hamiltonian for the one-dimensional unitary matrix model.
The constraint in \WWdeq\ reduces the space of states to singlets[\APA].
Hence, the problem is reducible to the eigenvalues of $W$.

Upon quantization, this problem is equivalent to a system of $N$
nonrelativistic fermions living on a circle, with the Hamiltonian given by
$$H=-\left({g^2L\over2}\right)\sum_{i=1}^N{\partial^2\over\partial\theta_i^2},
\qquad\qquad 0\le\theta_i<2\pi.\eqn\Hamferm$$
The fermionization is due to the Jacobian of the change of variables from
$W$ to its eigenvalues, introducing the Vandermonde-type
determinant in the wavefunction of the states, which in the unitary
matrix case reads
$$ {\widetilde \Delta} = \prod_{i<j} \sin{\theta_i - \theta_j \over 2}
= {\Delta (e^{i\theta_i}) \over \prod_i e^{i(N-1) \theta_i}},\eqn\VM$$
where $\Delta(\lambda_i ) = \prod_{i<j} (\lambda_i - \lambda_j )$ is the
standard Vandermonde determinant.
Notice that each factor in \VM\ is antiperiodic on the circle.
Thus, if $N$ is even the fermions have antiperiodic boundary conditions.
Likewise, if $N$ is odd they have periodic boundary conditions.
This can be understood in terms of transporting a fermion once around the
circle, passing by $N-1$ other fermions along the way
and therefore picking up $N-1$ minus signs.
Hence, in either case, the ground state is built by filling all states
with wave numbers between $-N/2+1/2$ and $N/2-1/2$, inclusive.
Subtracting off the ground state energy, one easily sees that this spectrum
reproduces that found for the different representations of $U(N)$.

If the gauge group is $SU(N)$, because $A_1$ is now traceless
$W$ will also obey the condition $\det W = 1$. Therefore the center of
mass coordinate for the fermions is absent and we must mod it out
of the theory. This means that we need to identify states in which all
fermions have their momentum shifted by the same amount.
Moreover, we must subtract the energy of the center of mass from the
energy of each state in the theory.

Now consider the partition function for the sphere.
In terms of the fermions, we want a matrix element that
corresponds to the sphere topology.  The obvious thing to do is
to map the end points of a cylinder to two points and therefore $W=1$ there.
In terms of the
fermions, this corresponds to computing the inner product of $N$ fermions
whose position is at the point $x=0$ on the circle at time $t=0$ with
the fermions at the same point at time $t=T$.  This inner product
$Z$ is given by
$$Z=\langle x_i ;t=T|y_i ;t=0\rangle,\eqn\matrxelm$$
in the limit that all $x_i\to 0$ and $y_i\to 0$.

To calculate $Z$, one can insert a complete set of momentum states for
each particle.  The wave function at $t=0$ and $t=T$ must be antisymmetric
under the exchange $x_i\to x_j$ or $y_i\to y_j$.  Therefore, the matrix element
in \matrxelm\ has the factor
$$\det|e^{ip_ix_j}|\det|e^{-ip_iy_j}|.\eqn\wffact$$
As $x_i\to 0$ and $y_i\to 0$, the factor in \wffact\ approaches
$$\prod_{i<j}(x_i-x_j)(y_i-y_j)(p_i-p_j)^2,
\eqn\wffactII$$
The matrix element in \matrxelm\ is therefore given by
$$Z=C \sum_{p_i}
\prod_{i>j}(p_i-p_j)^2(x_i-x_j)(y_i-y_j)
\exp(-\half g^2LT\sum_i p_i^2),\eqn\Zreduce$$
where $C$ is an unimportant constant.
Not surprisingly, this term approaches zero in this limit.
But in order to find the sphere contribution, one should notice that
the fermion wavefunction at the end points are {\it more}
singular than a $\delta$-function, namely
$$\psi (x_i ) = \widetilde\Delta (x_i ) \delta (W-1) = {1 \over \Delta (x_i )}
\delta (x_i ) \eqn\Sing$$
where a factor of $1/\widetilde\Delta^2 (x_i )$ was produced by the change of
variables from $W$ to $x_i$.
Therefore, it is necessary to divide the expression in \Zreduce\ by these
extra Vandermonde determinants, that is,
$$\prod(x_i-x_j)(y_i-y_j),$$
leaving a finite expression.  Hence the sphere partition function is
$$Z_{\rm sphere}=C\sum_{p_i}
\prod_{i>j}(p_i-p_j)^2 \exp(-\half g^2LT\sum_i p_i^2).\eqn\Zsphere$$

We can compare this to the sphere partition function of Migdal and
Rusakov[\Migdal,\Rus], which is given by
$$Z_{MR}=\sum_R (d_R)^2 \exp(-g^2A\CR)\eqn\MRsph$$
where the sum is over all representations of $U(N)$ or $SU(N)$ and
$d_R$ is the dimension of the representation and $\CR$ is the
quadratic casimir.
The correspondence of the fermion states with the $U(N)$ represenatations
is as follows[\Mike]:
If we describe a representation by a Young tableau, then the number
of boxes in row $i$, $n_i$ is the momentum shift
of the fermion with the $i^{\rm th}$ highest momentum
above its ground state value.
In terms of boxes, the casimir is given by
$$\CR=\half\left(N\sum_i n_i+\sum_i n_i(n_i-2i+1)\right)\eqn\casimir$$
which one can easily checked is reproduced by the fermions after
subtracting off the ground state energy.  The dimension of the
representation is given by
$$\eqalign{d_R&=\prod_{i>j}\left(1-{n_i-n_j\over i-j}\right)\cr
&=\prod_{i>j}(i-j)^{-1}\prod_{i>j}\Bigl((n_j-j)-(n_i-i)\Bigr).}\eqn\dimrep$$
The first product in the second line of \dimrep\ is a representation
independent term and is thus an unimportant
constant.  The second term is just $p_j-p_i$ for the fermions.
The total momentum is the $U(1)$ charge for a representation.
Hence we find full agreement with the result of Migdal and Rusakov.

\chapter{Classical Solutions}

After a cursory inspection of \Zsphere\ it would appear that $Z_{\rm sphere}$
is simply the partition function for a $d=0$ matrix model in a
quadratic potential. Unlike the lattice case, the potential never turns over,
so one might not expect a phase transition.  As was shown by Douglas and
Kazakov, this is not correct.  The point is that the variables $p_i$ that
appear in \Zsphere\ are discrete, hence the density of eigenvalues will
be bounded.  When this bound is reached, a phase transition occurs. Thus,
in the strong coupling phase we simply have condensation of the fermions
in their momentum lattice, which gives a very simple physical understanding
of the phase transition mechanism. The critical value of the area is reached
when the density of $p_i$, as given by the Wigner semicircle law which is
valid in the continuoum, reaches somewhere the lattice bound, namely one,
thus reproducing the DK result.

There should be a classical field configuration, termed master field, which
dominates the path-integral in each phase. In fact, as was shown by
Witten[\Witten]
using the localization theorem of Duistermaat and Heckman (DH),
the full \QCD\ path integral can be
written as a suitable sum over classical saddle points. In what follows we
will give a very simple demonstration of Witten's result using the fermion
picture and identify the classical configuration which dominates, that is,
the master field.

The key observation is that the exact propagator of a free particle is
proportional to the exponential of the action corresponding to the classical
(straight) path connecting the initial and final points. Since \QCD\ on the
torus is equivalent to $N$ free fermions, its partition function will also
be given by an appropriate classical path of the free fermions. The things to
be taken into account, however, are

i) The particles live on a circle; therefore there are several possible
classical paths for each of them, differing by their winding around the
circle with fixed initial and final positions.

ii) The particles are fermions; thus one should also consider paths where
the final positions of the particles have been permuted, weighted by a
fermionic factor $(-)^C$, where $C$ is the number of times the paths of
the particles cross.

The total partition function will then be the (weighted) sum of the actions
of all these classical configurations. Since to each
path corresponds a (diagonal) matrix $W(t)$, and to that (up to gauge
transformations) a classical field configuration satisfying the field
equations of motion, this is the sum over saddle points of the action
of Witten.

For the sphere the same picture holds, with the difference that all paths
start and end at the point $x=0$, and that each path is further weighted
by an extra factor, due to the division by the Vandermonde determinants as
explained in the previous section. This is, again, the sum over saddle
points of Witten, and the extra weighting factors are the determinants
which appear in the DH theorem. These paths are characterized by their
winding numbers $\{ n_i \}$ (up to permutation) and thus this is a sum of
the form
$$ Z_{\rm sphere} = \sum_{n_i} w( n_i ) \exp \Big(-{2\pi^2 \over g^2 LT}
\sum_i n_i^2 \Big)\eqn\Zcl$$
where $w( n_i )$ are the (as yet undetermined) weighting factors.

The easiest way to obtain the full expression in \Zcl\ is to Poisson resum
\Zsphere. Using the formula
$$ \sum_n f(n) = \sum_n {\tilde f} (2\pi n)\eqn\resum$$
where $f(x)$ is any function and $\tilde f$ is its Fourier
transform, we obtain
$$Z_{\rm sphere} = C \sum_{n_i} F_2 (2\pi n_i ),\eqn\Zres$$
where
$$ F_2 ( x_i ) = \int \prod_i dp_i e^{-i \sum_i x_i p_i } \Delta^2 ( p_i )
\exp(-\half g^2LT\sum_i p_i^2).\eqn\FF$$
To find the Fourier transform appearing in \FF, we first note that
$$ F_1 \equiv \int \prod_i dp_i e^{-i \sum_i x_i p_i } \Delta ( p_i )
\exp(-\half \alpha \sum_i p_i^2) = C \Delta ( x_i )
\exp(-{1 \over 2\alpha} \sum_i x_i^2) .\eqn\F$$
To prove this, notice that
$$ F_1 = \Delta ( -\partial_{p_i}) \int \prod_i dp_i e^{-i \sum_i x_i p_i }
\exp(-\half \alpha \sum_i p_i^2) = P ( x_i )
\exp(-{1 \over 2\alpha} \sum_i x_i^2).\eqn\FP$$
$P( x_i )$ is a polynomial of degree $N(N-1)/2$; moreover it is completely
antisymmetric in $x_i$. Therefore, up to a normalization, it is the
Vandermonde. The constant $C$ in \F\ can be found explicitly, it is however
irrelevant for this discussion since it will amount to an overall coefficient
in the final result. Using the convolution property of the Fourier transform
of a product, in combination with \F, we find
$$\eqalign{ F_2 ( x_i ) &= (F_1 \otimes F_1 ) ( x_i )\cr
& = C \int \prod_i dy_i
\Delta ({x_i - y_i \over 2}) \Delta ({x_i + y_i \over 2})
\exp\Big(-{1 \over 4g^2LT}\sum_i [(x_i+y_i)^2 + (x_i-y_i)^2 ]\Big)\cr
&= C \exp({1 \over 2g^2LT} \sum_i x_i^2 ) \int \prod_i dy_i \prod_{i<j}
(y_{ij}^2 - x_{ij}^2 ) \exp \Big(-{1 \over 2g^2LT}\sum_i y_i^2 \Big)}
\eqn\FFF$$
where $x_{ij}=x_i - x_j$ and $y_{ij}=y_i-y_j$. Substituting \FFF\ in
\Zres\ we recover expression \Zcl, with the weights $w( n_i )$ given by
$$ w( n_i ) = C \int \prod_i dy_i \prod_{i<j} (y_{ij}^2 - n_{ij}^2 ) \exp
\Big(-{1 \over 2g^2LT}\sum_i y_i^2 \Big) .\eqn\weights$$
The above determines the expression of the sphere partition function in
terms of classical saddle point configurations.

In the large $N$ limit one particular classical configuration in \Zcl\
should dominate. To determine this, let us ignore for the moment the fact
that the $n_i$ are discrete and replace the sum in \Zcl\ with an
integral. Then we can bring the expression for $Z_{\rm sphere}$ into the form
$$ Z_{\rm sphere} = C \int \prod_i ds_i dt_i \Delta (s_i) \Delta (t_i)
\exp \Big(-\half N \sum_i ( s_i^2 + t_i^2 )\Big),\eqn\Zlarge$$
where we rescaled $g^2$ to $g^2 /N$, in order to have a nontrivial large $N$
limit, and changed variables to $s=(y+2\pi n)/\sqrt{2g^2 LT}$,
$t=(y-2\pi n)/\sqrt{2 g^2 LT}$.
We see that in \Zlarge\ all explicit dependence on the coupling constant and
area have disappeared. The area enters this picture indirectly, through the
discreteness of $n_i$. (In fact, if it were not for this discreteness the
above integral would vanish.) Due to this, the plane $(s,t)$ is discrete in
the $s-t$ direction and consists of parallel diagonal lines with distance
$D=2\pi /\sqrt{g^2 LT}$. If this spacing is such that the distribution of
$s$ and $t$ contributing to the above integral in the large $N$ limit is
entirely within the lines $n=1$ and $n=-1$, then only the sector $n=0$
contributes. This signals a phase transition. Unlike the quantum
case, there is no exclusion principle for the windings $n_i$ and thus no
maximum density to be saturated. The phase transition in the classical
expansion is more like a Bose condensation to the ground state $n =0$.

To estimate the critical area, substitute the Vandermondes in \Zlarge\
with their absolute value, which corresponds to putting the product of
differences appearing in \weights\ into absolute values. This has no effect
for the $n_i=0$ term, while it overestimates the contribution of the other
sectors. \Zlarge\ then becomes the product of two independent
integrals involving a gaussian factor and one power of the Vandermonde.
The large $N$ saddle point of these integrals is found in a standard way
and the distribution of $s_i$ and $t_i$ is, in fact, again a Wigner
semicircle with radius $R=\sqrt2$. This on the $(s,t)$ plane creates a
``square" distribution with side $2R$ in each direction. When this square
lies entirely within the lines $n=\pm 1$, the sectors $n\neq 0$ do not
contribute and we are in the Boson condensate phase. For an estimate of the
critical area, assume that the transition occurs when the corners of the
square start touching the lines $n=\pm 1$, that is, when the half-diagonal
becomes $D$. Putting $2R\sqrt2 /2 = D$ we obtain $g^2 A_{\rm crit}=\pi^2$.
Somewhat surprisingly, our estimation has given the precise result of
Douglas and Kazakov. It is not entirely clear from this argument why
the transition should occur precisely at this point.

In any event, we see that the classical configuration in the weak-coupling
(small area)
phase is $n_i=0$. This corresponds to the master field $A=0$, up to
gauge transformations. As a check, notice that in this phase the discreteness
of $p_i$ in the quantum expression is irrelevant (no saturation) and thus
we can substitute the sums in \Zsphere\ by integrals. This, upon resummation,
becomes the $n_i=0$ term in \Zcl, as can be seen by the expression for $w(0)$
obtained from \weights. Only in the strong coupling phase will we have a
nontrivial master field. This is implicitly determined by \Zcl\ and \weights,
although its explicit expression is not known.

\chapter{$U(1)$ Sectors}

There is one difficulty with the string picture as it now stands for
the $U(N)$ gauge group, namely, the sum is not asymptotic with the exact
answer in the limit $1/N\to0$.
This is because there exist states with finite energy, but which will never
appear in the perturbative sum in \MRsph.
These are the states that correspond to $N$ fermions with momenta
shifted by a constant finite amount.  Such states have finite energy in the
large $N$ limit, but do not show up in a perturbative
sum over surfaces since the corresponding Young tableau has at least
$N$ boxes.  Moreover, these states are local minima, in the sense that in order
to find a state with lower energy, it is necessary to shift the total momentum
by a large amount.

We can rectify this problem by including other sectors in the sum.  For finite
$N$, this will eventually lead to overcounting, but the answer will be
asymptotic.  To this end, let us define $Z_m$
$$Z_m=\sum_{\rm Reps}d_R^2\exp\left(-{Ag^2\over2}\sum_i(n_i+m)\right)
\exp\left(-{Ag^2\over2N}\sum_i(n_i-2i+1+m)(n_i+m)\right),\eqn\Zmeq$$
where we have basically added $m$ boxes to each row.  After some manipulation,
$Z_m$ reduces to
$$Z_m=\sum_{\rm Reps}d_R^2e^{-{Ag^2\over2}(1+2m/N)n_R}
e^{-{Ag^2\over2N}\nt_R}e^{-Ag^2m^2\over2},
\eqn\ZmeqII$$
Hence we see that $Z_m$ has the same form as $Z_0$, except that
the area term that appears in front of $n$ has been shifted and there is
an extra factor of $m^2$ in the energy.
But the $\nt$ term is the same, meaning that
the Gross-Taylor rules are exactly the same as in the $Z_0$ case,
except that the area that will be used for the Nambu-Goto term has
been renormalized, with a different shift for the chiral and antichiral
sectors.
The sum
$$\cZ=\sum_mZ_m\eqn\Zstring$$
is now an asymptotic sum for the \QCD\ string.

We can interpret the sum in \Zstring\ as a sum over different
classical string solutions, since around each sector $m$ there is a
sum over $1/N$, the string coupling.   These different sectors are
basically the conjugates of the $U(1)$ instanton sectors.
We can Poisson resum \Zstring, giving the expression
$$\cZ=\sqrt{\pi\over Ag^2}\sum_m\sum_{\rm Reps}d_R^2
e^{-({Ag^2\over2}+{2\pi im\over N})n_r} e^{-{Ag^2\over2N}\nt_R}
e^{Ag^2n_R^2\over2N^2}
e^{-{m^2\over2g^2A}}.\eqn\Zprsm$$
The $m^2$ term now has a factor of $1/g^2$ in front of it, which is the
contribution one would expect from $U(1)$ instantons.  Furthermore, the
area that now appears in the Nambu-Goto action is now complex\foot{
This might be useful for formulating \QCD\ as a topological field
theory.  We thank S. Cordes and C. Vafa for informing us of their work
in this area.}, and there
is also an additional $n^2$ term.  This last term has been discussed before
in the $SU(N)$ case and was attributed to contributions from tubes and handles
on the \ws.

Since there are other string solutions corresponding to different values
of $m$, we might expect to be able to generalize the work of Douglas
and Kazakov to find the leading order contributions around these solutions.
To this end, consider the solutions to the equations of motion for
the path integral in \Zsphere,
$${Ap_i\over2N}=\sum_{j\ne i}{1\over p_i-p_j},\eqn\eqmdis$$
where we have absorbed the QCD coupling into the area and rescaled it
by a factor of $N$.  In the large $N$ limit we can define the new
variables
$x=i/N$, and $h(x)=p_i/N$, giving the new equation,
$$Ah/2={\rm P}\int d\lam {u(\lam)\over h-\lam},\eqn\eqmcont$$
where $u(\lam)$ is the density of eigenvalues, $u(\lam)=dx/d\lam$.
But unlike the eigenvalues of a matrix model in a quadratic potential,
the $p_i$ are discrete, therefore the density $u(\lam)$ is bounded,
satisfying $u(\lam)\le1$.

DK treated this problem
by dividing up the possible real values of $\lam$ into three regions,
where $u(\lam)=0$, $0<u(\lam)<1$ and $u(\lam)=1$.  They choose
the ansatz that the first region
occurs for $|\lam|>a$, the second for $b<|\lam|<a$, and the third
for $|\lam|<b$.  If we define $\tu(\lam)$ to be the density of
eigenvalues minus the contribution from the region where $u(\lam)=1$,
then \eqmcont\ can be rewritten as
$$Ah/2+\log{h-b\over h+b}={\rm P}\int d\lam{\tu(\lam)\over h-\lam},
\eqn\eqmtcont$$
The problem has been reduced to a two cut eigenvalue problem.  To solve
this, define the function
$$f(h)=\int d\lam{\tu(\lam)\over h-\lam}.\eqn\feq$$
$f(h)$ then must satisfy
$$f(h)={1\over 2\pi i}\sqrt{(h^2-a^2)(h^2-b^2)}
\oint ds {As/2-\log{b-s\over b+s}\over (h-s)\sqrt{(a^2-s^2)(b^2-s^2)}},
\eqn\fcontour$$
where the contour surrounds the cuts from the square roots, but not
the cut from the log nor the singularity at $h$.
Pulling the contour back, \fcontour\ leads to the equation
$$f(h)=h{A\over2}+\log{h-b\over h+b}-\sqrt{(a^2-h^2)(b^2-h^2)}
\int_{-b}^b ds {1\over(h-s)\sqrt{(a^2-s^2)(b^2-s^2)}}.\eqn\feqcon$$

Expanding about large $h$, one can then find relations between $a$,
$b$ and $A$.  Before doing this, let us go back and reexamine
the DK ansatz.  They of course make the reasonable
assumption that the solution to the equations of motion is symmetric
about $h=0$.  This is quite sensible for the ground state, but should
not be true for solutions that correspond to sectors with nonzero
values of $m$.  Therefore, let us relax their ansatz somewhat and
assume that $0<u(\lam)<1$ for the regions $b<\lam<a$ and $d<\lam<c$,
where $a$, $b$, $c$ and $d$ are to be determined.  Then we can then
proceed as before, reaching the equation
$$f(h)=h{A\over2}+\log{h-b\over h-c}-
\int_{c}^b ds {\sqrt{(a-h)(b-h)(c-h)(d-h)}
\over(h-s)\sqrt{(a-s)(b-s)(s-c)(s-d)}}.\eqn\feqnew$$
{}From \feq\ and (A.6) one finds that the density of eigenvalues is given by
$$\eqalign{u(\lam)&={\sqrt{(a-\lam)(\lam-b)(\lam-c)(\lam-d)}\over\pi}\int_c^b
ds{1\over(\lam-s)\sqrt{(a-s)(b-s)(s-c)(s-d)}}\cr
&={2\over\pi(\lam-c)(\lam-d)}
{\sqrt{(a-\lam)(\lam-b)(\lam-c)(\lam-d)}\over\sqrt{(a-c)(b-d)}}\cr
&\qquad\qquad\qquad\times
\left((c-d)\Pi({b-c\over b-d}{\lam-d\over \lam-c},q)+(\lam-c)K(q)\right),}
\eqn\density$$
where $K$ and $\Pi$ are the complete elliptic integrals of the first
and third kind.

Let us now expand the righthand side \feqnew\ in powers of $1/h$ and compare
it to the expansion in \feq.  Matching the terms of order $h$ gives
the equation
$${A\over2}-\int_c^b ds {1\over\sqrt{(a-s)(b-s)(s-c)(s-d)}}=0.\eqn\hexpI$$
Using equation (A.1) in the appendix gives
$$A={4K(q)\over\rho},\eqn\hIre$$
where
$$\rho=\sqrt{(a-c)(b-d)},\qquad\qquad q=\sqrt{(a-d)(b-c)\over(a-c)(b-d)}.$$
Note that for the case $c=-b$, $a=-d$, the modulus $q$ that appears
in \hIre\ is different than the corresponding equation in [\DK].

Matching the $h^0$ terms in the expansion of \feq\ and \feqnew\ leads
to the equation
$$\half\int_c^b ds {a+b+c+d\over\sqrt{(a-s)(b-s)(s-c)(s-d)}}=
\int_c^b ds {s\over\sqrt{(a-s)(b-s)(s-c)(s-d)}}.\eqn\hexpII$$
Using (A.1) and (A.2), we can rewrite this equation as
$${c-d\over\rho}\Pi(\al,q)={a+b+c-d\over2\rho}K(q),\eqn\hIIre$$
where $\al=\sqrt{b-c\over b-d}$.  In [\DK], the $h^0$ equation is
trivially satisfied due to symmetry.

Matching the $h^{-1}$ term in the two equations gives
$$\eqalign{\int ds \tu(s)+b-c&=-\int_c^b
ds{s^2\over\sqrt{(a-s)(b-s)(s-c)(s-d)}}
\cr
&+\half\int_c^b ds {s(a+b+c+d)\over\sqrt{(a-s)(b-s)(s-c)(s-d)}}\cr
&+{1\over8} \int_c^b ds {a^2+b^2+c^2+d^2-2ab-2ac-2ad-2bc-2bd-2cd
\over\sqrt{(a-s)(b-s)(s-c)(s-d)}}.}\eqn\hexpIII$$
The integral on the lefthand side of \hexpIII\ is the total number
of eigenvalues minus those in the region between $c$ and $b$, divided
by $N$.  But the number between $c$ and $b$ is just $N(b-c)$, thus the
lefthand side is unity.
Using (A.1), (A.2) and (A.3), and then using the equation
in \hIIre, we can reduce \hexpIII\ to
$$1={(a-b-c+d)^2\over4\rho}K(q)+\rho E(q),\eqn\hIIIre$$
where $E(q)$ is the complete elliptic integral of the second kind.

The next term in the expansion in \feq\ is
$$h^{-2}\int ds \tu(s)s.$$
But this is just the sum of the momenta for this particular solution
coming from the regions where the density is less than one,
divided by $N^2$.  This is its contribution to the $U(1)$ charge divided by
$N^2$.   Let us call this rescaled charge $\tQ$.
Using equations (A.1)-(A.4), \hIIre\ and \hIIIre, the $h^{-2}$
expansion leads to the equation
$$\tQ+\half(b^2-c^2)=Q={W\over4}+{K(q)\over16}XYZ,\eqn\hIVre$$
where
$$\eqalign{W&=a+b+c+d\qquad\qquad\qquad X=a-b-c+d\cr
           Y&=a+b-c-d\qquad\qquad\qquad Z=a-b+c-d.}\eqn\neweqs$$
$(b^2-c^2)/2$ is the contribution to the rescaled charge from the
eigenvalues that sit between $c$ and $b$, hence the lefthand side of
\hIVre\ is the total rescaled charge.  The charge for the lowest
energy state in sector $m$ is $mN$, hence $Q=m/N$.

Finally, we can find the specific heat, which is the next term in the expansion
of $f(h)$.
Using (A.1)-(A.5), \hIIre\ and \hIIIre, we find,
after a fair amount of algebraic manipulation
$$\eqalign{F'(A,Q)-1/24&=\int ds u(s)s^2\cr
&={\rho E(q)\over48}(3W^2+X^2+Y^2+Z^2)\cr
&\qquad+{K(q)\over12\rho}
(3W^2X^2+6WXYZ+Y^2Z^2+2X^2Y^2+2X^2Z^2+X^4)\cr
&={K(q)\over12\rho}
(6WXYZ+Y^2Z^2+X^2Y^2+X^2Z^2)\cr
&\qquad+{1\over48}(3W^2+X^2+Y^2+Z^2).}\eqn\spheat$$

Given the area $A$ and the charge $Q$,
in principle, one should be able to
solve for $a$, $b$, $c$ and $d$ using the four equations \hIre, \hIIre,
\hIIIre\ and \hIVre.  However, we should not expect solutions for all
possible values of $A$ and $Q$.  For instance, from DK, we know that
there are no solutions for $Q=0$ and $A<\pi^2$.
This of course is the weak coupling regime.  But for nonzero values
of $Q$, the minimum value of $A$ should be higher.  In fact, in the
infinite area limit, the upper value of $Q$ is bounded.  Naively, this
should happen for the sector when the fermion with the smallest or largest
momentum is zero. We will see that the naive answer is correct.

To further analyze the problem, we will consider two regions.  The first
corresponds to the area near its critical value for small values of $Q$.
The second region occurs for very large values of the area.
Let us first consider what happens at the critical point.  In this case
$b=c$, thus $\al=q=0$ and $\Pi=K=E=\pi/2$. From equation \hIIre\ we
find that $a=-d$ at the critical point.
To determine $b$, let us go back to equation \density.  Since $b=c$, we
find that $u(\lam)$ is given by
$$\eqalign{u(\lam)&={\sqrt{a-\lam\over a+\lam}}{1\over\sqrt{a^2-b^2}}
[(\lam-b)+(b+a)]\cr
&=\sqrt{a^2-\lam^2\over a^2-b^2}.}\eqn\dencrit$$
If $b\ne0$, then for $\lam^2<b^2$, $u(\lam)>1$.  But this violates the
ansatz that the density is less than or equal to unity.  Therefore,
we must choose $b=0$.  Using \hIIIre\ and \hIVre\ we then see that
$a={2\over\pi}$ and that the charge for this solution is zero.

Now consider small, but nonzero values for $Q$.  In this case,
$b-c$ must be nonzero.  To this end, let
$$\eps=b-c,\qquad\qquad\de=b+c\eqn\bpc$$
Using \hIIIre\ and the asymptotic expansions in the appendix,
we find the leading correction to $a$ from its critical value is
$$\Delta a=-{\pi\over32}(\eps^2+2\de^2).\eqn\acorr$$
Since this correction is of order $\eps^2$ and not $\eps$,
it will not contribute to $a+d$ in leading order.
This leading order correction can be found from
\hIIre\ and the asymptotic expansions in the appendix.  A little algebra
shows that
$$a+d={\pi^2\over32}\eps^2\de.\eqn\apdcorr$$
We can now use \hIVre, \bpc\ and \apdcorr\ to find $Q$.
Let us rewrite \hIVre\ as
$$\eqalign{Q&=(a+d)\left({1\over4}+{K\over16\rho}(a-d+b-c)(a-d-b+c)\right)\cr
&\qquad\qquad
+(b+c)\left({1\over4}-{K\over16\rho}(a-d+b-c)(a-d-b+c)\right).}\eqn\chargere$$
The factor multiplying $a+d$ is to leading order $1/2$.  The term multiplying
$b+c$ is actually much smaller.   In fact,  this term is third order in $\eps$
and is given by $-\pi^3\eps^3/1024$.  Hence, the leading contribution to
the charge actually comes from the $a+d$ term and is
$$Q={\pi^2\over64}\eps^2\de.\eqn\chcorr$$

The charge that appears in \chcorr\ is limited by the maximum value of
$\de$ given $\eps$.  This bound is determined by enforcing the ansatz
that $u(\lam)\le1$.  The region where this ansatz might be violated
is where $\lam\approx b$ or $\lam\approx c$.  Let us
consider the case where $\lam$ is near $b$.  Examining equation \density,
we see that the first modulus in the elliptic integral of the third kind
approaches unity as $\lam$ approaches $b$ from above.
Therefore, if we substitute the asymptotic expansion for $\Pi$ in (A.11)
in \density, we find that the density is given by
$$\eqalign{u(\lam)&= {2\over\pi}{1\over b-d}\sqrt{(a-b)(\lam-b)\over(a-c)(b-c)}
\Biggl\{(b-c)K(q)+(c-d)K-E(q){(a-c)(b-d)\over a-b}\cr
&\qquad\qquad\qquad+(c-d){\pi\over2}\sqrt{(a-c)(b-d)\over(a-b)(c-d)}
\sqrt{(b-d)(b-c)\over(\lam-b)(c-d)}
\Biggr\}+{\rm O}(\lam-b)\cr
&=1+\sqrt{\lam-b}{2\over\pi}\sqrt{a-b\over(a-c)(b-c)}\left(K(q)-E(q){a-c\over
a-b}
\right)+{\rm O}(\lam-b).}\eqn\densityapp$$
Hence, in order to ensure that the ansatz is satisfied, it is necessary
that the relation
$${a-c\over a-b}E(q)-K(q)>0,\eqn\EKrel$$
be upheld.  Using the asymptotic expansions and the values for $b$, $c$, $a$
and $d$ given by \bpc\ and \acorr, we find that
$${a-c\over a-b}E(q)-K(q)\approx{\pi^3\over32}(2\de\eps+\eps^2).\eqn\epsde$$
Therefore, in order for \EKrel\ to be satisfied, we must have $\de>-\eps/2$.
We can also derive the constraint that $\de<\eps/2$ by examining $u(\lam)$
as $\lam\to c$.  Therefore, we must satisfy
$$|\de|<\eps/2.\eqn\delepscon$$
We can rewrite this constraint in terms of the charge and the area. From
\hIre\ and the asymptotic expansion, we have the relation
$$A-A_c={3\pi^4\over64}(\eps^2+2\de^2),\eqn\Aeps$$
where $A_c$ is the critical value for the area.
Hence, using \chcorr, \delepscon\ and \Aeps, we have that
the maximum allowed charge sector for a given area near its critical value is
$$Q_{max}={32\sqrt{2}\over27\pi^4}(A-A_c)^{3/2}.\eqn\Qmax$$
This relation is a little reminiscent of the maximum charge of a
Reissner-Nordstrom black hole.

Now we wish to examine the behavior deep in the strong coupling regime,
which corresponds to large values of the area.
In this case, we expect $b$ to approach $a$ and $c$ to approach $d$.  The
values of $a$ and $d$ are determined by the charge of the sector that we
are considering.  For this behavior $q\to1$, and thus $q'\to0$.  At this
point it is convenient to rewrite $\Pi(\al,q)$ in terms of elliptic
integrals of the first and second kind.  Using the relation in the
appendix, equation \hIIre\ can be written as
$$-E(q)F(\th,q)+K(q)E(\th,q)={a+b-c+d\over2\rho}K(q),\eqn\hIIinc$$
where $\sin\th={a-c\over a-d}$
Using \hIIIre\ and the asymptotic expansions in (A.12) and (A.13),
we find that in this region
$$a-d\approx1\eqn\adrel$$
which is expected since the the integral of $u(\lam)$ should be unity
and for almost all values of $\lam$ between $a$ and $d$, $u(\lam)=1$.
Plugging in leading order asymptotic expansions in (A.12)-(A.15)
into \hIIinc\ and invoking \adrel, we then find the following approximate
equation
$$-\log{1+\sin\th\over\cos\th}+\log\sqrt{(a-c)(b-d)\over(a-b)(c-d)}\to
\log\sqrt{1\over a-b}\approx a\log\sqrt{1\over(a-b)(c-d)}.\eqn\logrel$$
If we let $a-b=\eps$ and $c-d=\eps^\mu$, then \logrel\ gives
$$a={1\over1+\mu}.\eqn\aeq$$
Since $0<\mu<\infty$, we see that the possible values of $a$ range from
0 to 1.  Of course what this means is that no local solutions exist
if the fermions are shifted such that all of the momenta are greater
than 0 or all are less than 0.  This should not come as a big surprise,
since once these limits are reached, then there are small deformations
of the fermion momenta which lower the energy of the state.

The charge for these solutions is dominated by the $W$ term in \hIVre,
since $X$ $Y$ and $Z$ are all small.  Clearly in the limit $b\to a$ and
$c\to d$, the charge approaches
$$Q\to {a+d\over2}=a-\half.\eqn\Qlarge$$

\chapter{Correspondence with Lattice Models}

Douglas and Kazakov have remarked that the phase transition for the continuum
model is similar to the phase transition that occurs for the lattice.  In
both cases the phase transition is third order and the equations of motion
for the eigenvalues are given by a two cut model.  In some sense, these
two situations are dual to each other, with the weak coupling region of the
lattice model acting like the strong coupling region of the continuum
case.

However, it would appear that this correspondence breaks down when we
consider the solutions with nonzero values of $Q$.  There are no
corresponding solutions for weakly coupled $U(N)$ on the lattice.  But
a little more thought shows that correspondence is between continuum $U(N)$
and the lattice $SU(N)$ and vice versa.  For instance, the continuum
$SU(N)$ case does not have these extra solutions, since the $U(1)$ charge
is not a degree of freedom.  In terms of the fermions, the center of mass
coordinate is modded out.  Hence shifting all the fermion momenta by
the same amount gives back the same state.  On the other hand, the $SU(N)$
case has an extra term in the casimir, $n^2/N^2$, where $n$ is the number
of boxes in the representation. But this term does not survive the scaling
limit, so we can safely drop it.

For $SU(N)$ on the lattice, the eigenvalues sit in a potential described
by \matpot.  However, unlike the $U(N)$ case, the center of mass position
for these eigenvalues must satisfy the constraint that
$$\sum_i\th_i=2\pi m,\eqn\thcon$$
where $m$ is an integer.  Clearly, the $U(N)$ classical solution is the
same as the lowest energy state for the $SU(N)$ case, which will have
$m=0$.  But suppose we consider a case where $0<m<N/2$.  Since $m$ is
an integer, we cannot smoothly deform this to the $m=0$ classical
solution.  Hence each value of $m$ must have a classical minimum.  We
could also have values of $m$ which are less than 0, but shifting the
total position by $2\pi N$ just shifts all eigenvalues around the
circle, hence these correspond to the same solution.
The maximum value of $m$ for a given weak coupling is determined by
the value which puts an eigenvalue at the top of the potential. Clearly as
the coupling becomes stronger, the number of classical solutions will
decrease.

Let us discuss this situation in  a little more detail.  We can impose the
constraint in \thcon\ by inserting a $\de$-function into the path integral
with the form
$$\de(\sum_i\th_i-2\pi m)={\lim\atop\eps\to0}{1\over\sqrt{2\pi\eps}}
e^{-{1\over2\eps}(\sum_i\th_i-2\pi m)^2}.\eqn\thdf$$
Hence, using \Vpot\ we find the equation of motion
$$N\be\sin\th_i-\sum_{j\ne i}\cot{\th_i-\th_j\over2}
-{1\over\eps}(\sum_i\th_i-2\pi m)=0.\eqn\eqmot$$
where $\be=1/(a^2g^2)$.
Under the usual rescaling, one ends up with the equation
$$\be\sin\th-{1\over\eps}(\int\mu(\phi)\phi-2\pi m/N)={\rm P}\int\mu(\phi)
\cot{\th-\phi\over2},\eqn\eqscaled$$
where $\mu(\phi)$ is the density of eigenvalues.  As $\eps\to0$, the last
term on the lhs of \eqscaled\ is divergent unless the integral is very close
to $2\pi m$.  To this end, let us replace this entire term by a constant
$\al$.  Next define the function $f(\th)$,
$$f(\th)=\int d\phi \mu(\phi) \cot {\th-\phi\over2}.\eqn\fth$$
Given this definition, the function is analytic everywhere in the strip
$-\pi<\re\th<\pi$, except for a cut along the real line.  $f(\th)$ is
clearly invariant under $\th\to\th+2\pi$.

$f(\th)$ can be solved for
by analyzing its behavior as $\im\th\to\infty$.  In particular, using \fth\
the large $\im\th$ behavior for $f(\th)$ is
$$f(\th)=-i\int d\phi\mu(\phi)-ie^{i\th}\int d\phi e^{-i\phi}\mu(\phi)+
{\rm O}(e^{2i\th}).\eqn\fthas$$
Let us assume that the eigenvalues sit on the strip $\De-a<\lam<\De+a$.  Then
from \eqscaled\ and \fth, $f(\lam\pm i\eps)$ satisfies
$$f(\lam\pm i\eps)=\be\sin(\lam)+\al\mp2i\pi\mu(\lam).\eqn\fcut$$
Choosing the periodic function
$$f(\th)=\be\sin\th+\al-2\be\cos{\th+\De\over2}\sqrt{\sin^2{\th-\De\over2}-
\sin^2{a\over2}},\eqn\ffunc$$
one finds that the asymptotic behavior is matched if
$$\be\sin^2{a\over2}\cos\De=1,\eqn\asyone$$
and
$$\be\cos^2{a\over2}\sin\De=-\al.\eqn\asytwo$$
{}From \fcut\ and \ffunc\ one learns that the density of eigenvalues is given
by
$$\mu(\lam)={\be\over\pi}\cos{\lam+\De\over2}
\sqrt{\sin^2{a\over2}-\sin^2{\lam-\De\over2}}.\eqn\eigden$$
In order that the density of states is positive, one must satisfy
the inequalities $\De+a\le\pi$ and $\De-a\ge-\pi$.  One can also easily show
that the integral of $\sin\phi$ weighted by the density of eigenvalues
satisfies
$$\int\mu(\phi)\sin\phi=\cos^2{a\over2}\sin\De.\eqn\sinwt$$
Hence $\De$ basically measures the anisotropy of the solution about the
bottom of the well.

As $\be\to\infty$, it is clear that the allowed solutions for $\De$ lie
anywhere between $-\pi/2$ and $\pi/2$.  This is similar to the situation
in the previous section, where any charge between $-N/2$ and $N/2$ is
allowed as the area approaches infinity.    However, near the critical point,
only a small range for $\De$ is allowed.  Clearly, from \asyone, the critical
value of $\be$ is $\be_c=1$.  Therefore, the only
allowed $\De$ at the critical point is $\De=0$.

Moving slightly away from the critical point, one has $a=\pi-\eps$ and a
maximum value of $\De$ given by $\De=\eps$.  At the maximum, the integral
of the angles weighted by the density of eigenvalues is then given by
$$\eqalign{
\int d\phi\mu(\phi)\phi&={\be\over\pi}\int_{\De-a}^{\De+a}
d\phi \phi\cos{\phi+\De\over2}
\sqrt{\sin^2{a\over2}-\sin^2{\phi-\De\over2}}\cr
&=\De-{\be\over\pi}\sin\De\int_{-a}^{a}d\phi \phi\sin\phi\sqrt{\sin^2{a\over2}-
\sin^2{\phi\over2}},}\eqn\intth$$
where we have shifted the integration variables and have used the fact that
the integral over the density of states is one.  For small $\eps$ we find
that the leading order term in \intth\ is
$$
\int d\phi\mu(\phi)\phi={1\over8}\eps^2\De\log{1\over\eps}+{\rm O}(\eps^2\De).
\eqn\intlead$$
$\be$ is the inverse area, hence we find that the area for the plaquette
satisfies
$$A_c-A={1\over4}(\eps^2+2\De^2).\eqn\Alat$$
Therefore, the maximum charge in terms of the area, behaves
like
$$Q_{\rm max}\sim (A_c-A)^{3/2}\log{1\over A_c-A}.\eqn\Qmlat$$
Hence, unlike the continuum case, the maximum charge in terms of the
area has a scaling violating piece.
However, the integral over $\sin\phi$ has a maximum that behaves like
$(A_c-A)^{3/2}$, which is closer to the behavior of the preceeding section.

\chapter{Discussion}

In this paper we have given an alternative derivation of Witten's
proof that the QCD partition function is given by a sum over
saddle points.
We then showed how the sphere phase transition occurs in the
dual picture of QCD string theory.

We next demonstrated how to find new solutions for the $U(N)$
\QCD\ sphere partition function corresponding to the different $U(1)$
sectors.  We have also shown how these solutions relate to solutions
found for the lattice versions of $SU(N)$.
In some sense, it is surprising that the nonzero charge solutions can actually
be found.  One might have expected that since there is nothing in the
machinary of section 4 that explicitly states that the eigenvalues
lie on the lattice, then nothing should prevent the nonzero charge
solutions from sliding down to the absolute minimum.
But actually, this information is in there, because of the $\log$ term
that appears in \fcontour. This term states that there is a region
where the eigenvalues have a constant density, that is, they lie on
the lattice.  The boundaries of this region essentially add another
degree of freedom.  This then allows us to find solutions with nonzero
charges.

It is hoped that the results presented here might have some use in
investigating random matrix model theory.  Perhaps one can consider
cases where the entries of the Hamiltonian are restricted to be
integers.  One might then find similar behavior to that shown here.

\appendix

In this appendix we present some useful formulae of elliptic integrals.
These are taken from or are easily derived from formulae in [\Byrd].
The first such equations are
$$\int_c^b ds {1\over\sqrt{(a-s)(b-s)(s-c)(s-d)}}=
{2\over\rho}K(q),\eqn\ellipI$$
$$\int_c^b ds {s\over\sqrt{(a-s)(b-s)(s-c)(s-d)}}=
{2\over\rho}\Bigl((c-d)\Pi(\al,q)+dK(q)\Bigr)
,\eqn\ellipII$$
$$\eqalign{\int_c^b& ds {s^2\over\sqrt{(a-s)(b-s)(s-c)(s-d)}}=\cr
&{1\over\rho}\Bigl((-ad-bc+d(a+b+c+d))K(q)-\rho^2E(q)
+(a+b+c+d)(c-d)\Pi(\al,q) \Bigr),}\eqn\ellipIII$$
$$\eqalign{\int_c^b& ds {s^3\over\sqrt{(a-s)(b-s)(s-c)(s-d)}}=\cr
&{1\over\rho}\Bigl(2d^3-{1\over4}(c-d)(b-d)(a+3b+3c+5d)\Bigr)K(q)
-{3\over4}(a+b+c+d)\rho E(q)+\cr
&{1\over\rho}\biggl({3\over4}(a+b+c+d)(a+b+c-3d)\cr
&\qquad\qquad\qquad
+(2a+2b+2c+3d)d-ab-ac-bc\biggr)(c-d)\Pi(\al,q),}\eqn\ellipIV$$
and
$$\eqalign{\int_c^b& ds {s^4\over\sqrt{(a-s)(b-s)(s-c)(s-d)}}=\cr
&{1\over\rho}\Bigl(2d^3-{1\over24}(c-d)(b-d)[-5a^2-15b^2-15c^2-33d^2\cr
&\qquad\qquad\qquad-4ab-4ac-6ad-14bc-24bd-24cd]\Bigr)K(q)\cr
&-{1\over24}\Bigl(a5a^2+15b^2+15c^2+15d^2+14ab+14bc+14ac+14bc+14bd
+14cd\Bigr)\rho E(q)\cr
&+{1\over8\rho}\Bigl(5a^3+5b^3+5c^3+5d^3+3a^2b+3a^2cx+3ab^2
+3b^2c+3bc^2+3a^2d\cr
&\qquad\qquad\qquad+3ad^2+3bd^2+3cd^2+2abc+2abd+2acd+2bcd\Bigr)
(c-d)\Pi(\al,q),}\eqn\ellipV$$
where
$$\rho=\sqrt{(a-c)(b-d)},\qquad\qquad \al=\sqrt{{b-c\over b-d}},
\qquad\qquad q=\sqrt{{(a-d)(b-c)\over(a-c)(b-d)}}.\eqn\rhoalq$$
$K(q)$, $E(q)$ and $\Pi(\al,q)$ are the complete elliptic integrals
of the first, second and third kind respectively.

Another formula used in the text is
$$\eqalign{\int_c^b&{1\over(\lam-s)\sqrt{(a-s)(b-s)(s-c)(s-c)}}\cr
&={2\over(\lam-c)(\lam-d)\sqrt{(a-c)(b-d)}}
\Bigl[(c-d)\Pi({b-c\over b-d}{\lam-d\over \lam-c},q)+(h-c)K(q)\Bigr].
}\eqn\ellipden$$

We also use the relation between the complete integral of third kind
and incomplete integrals of the first and second kind,
$${(c-d)\over\rho}\Pi(\al,q)={(c-d)\over\rho}K(q)-E(q)F(\th,q)+K(q)E(\th,q),
\eqn\comptoinc$$
where $\sin^2\th={a-c\over a-d}$, and $F(\th,q)$ and $E(\th,q)$ are the
incomplete integrals of the first and second kind.

The following asymptotic expansions are also used:
$$K(q)={\pi\over2}\left(1+{1\over4}q^2+{9\over64}q^4+{25\over256}q^6+...\right)
,\eqn\Kasym$$
$$E(q)={\pi\over2}\left(1-{1\over4}q^2-{3\over64}q^4-{5\over256}q^6+...\right)
,\eqn\Easym$$
$$\eqalign{\Pi(\al,q)=&{\pi\over2}\biggl(1+{1\over2}(\al^2+q^2/2)
+{24\over64}(\al^4+ \al^2q^2/2+3q^4/8)\cr
&\qquad+{5\over16}(\al^6+\al^4q^2/2+3\al^2q^4/8+5q^6/16)+...\biggr)
,}\eqn\Piasym$$
$$K(q)=\log(4/q')-2[\log(4/q')+1](q')^2+...,\eqn\Kasymp$$
$$E(q)=1+{1\over2}[\log(4/q')-1/2](q')^2+...,\eqn\Easymp$$
$$F(\th,q)=\log{1+\sin\th\over\cos\th}+{1\over4}[\log{1+\sin\th\over\cos\th}
-\sin\th\sec^2\th](q')^2+....\eqn\Fasym$$
$$E(\th,q)=\sin\th+2[\log{1+\sin\th\over\cos\th}
-\sin\th](q')^2+....\eqn\Easym$$
where $q'$ satisfies $q'=\sqrt{1-q^2}$.

Finally, there is a useful expansion for $\Pi(\al,q)$ if $\al$ is close
to unity.  This is
$$\Pi(\al,q)=K(q)-{E(q)\over(q')^2}+{\pi(1+(q')^2-q^2\al^2)\over
4(q')^3\sqrt{1-\al^2}}+{\rm O}(1-\al^2).\eqn\Pialasym$$

\ack{The research of J.A.M.~was supported in part by D.O.E.~grant
DE-AS05-85ER-40518.  A.P.P.~acknowledges helpful conversations with
M.~Bauer. J.A.M.~does the same with M.~Douglas.}

{\it Note added:}
As this paper was being typed we received a paper[\CDMP] which discusses
some of the issues in section 3.

\refout
\end